\documentclass[a4paper,11pt]{article}
\usepackage{pos}

\title{Status of the MUonE experiment}

\author*[1]{Giovanni Abbiendi}

\affiliation{INFN - Sezione di Bologna,\\
Viale C. Berti Pichat, 6/2, 40127 Bologna  -   Italy}

\note{For the MUonE collaboration.}

\emailAdd{giovanni.abbiendi@cern.ch}

\abstract{
The MUonE experiment aims at an independent and very precise
determination of the leading hadronic contribution to the muon magnetic
moment, based on an alternative method, complementary to the existing ones.
This can be achieved by measuring with unprecedented precision
the shape of the differential cross section of $\mu e$ elastic
scattering, using the intense muon beam available at CERN, with energy
of 150 GeV, off atomic electrons of a light target.
The status of the project is presented, with recent results in
preparation for the test run scheduled in 2021 with a reduced detector.
}

\FullConference{%
  40th International Conference on High Energy physics - ICHEP2020\\
  July 28 - August 6, 2020\\
  Prague, Czech Republic (virtual meeting)
}

\begin{document}
\maketitle

\section{Introduction}
The muon magnetic moment is one of the most precisely measured
quantities (better than one part per billion).
It is also calculable with extremely high precision in the Standard
Model, hence it constitutes one of the most stringent tests of the
theory.
Currently the reference measurement \cite{Bennett:2006fi} shows a significant discrepancy of
$3.7\sigma$ with respect to the Standard Model prediction \cite{Aoyama:2020ynm}, which is
one of very few deviations in the high energy physics landscape.
A new measurement is underway at Fermilab \cite{Grange:2015fou}, targeting a reduction of the
experimental error by a factor of 4, and another one is planned at
J-PARC \cite{Abe:2019thb}.
In view of the forthcoming improved measurements a reduction of the
theoretical error is also desirable, in particular of its largest contribution,
related to the hadronic vacuum polarization, which is not calculable
in perturbation theory.
At leading order this is usually evaluated via a dispersion integral of the hadron production cross section in $e^+e^-$
annihilation \cite{Davier:2019can}.
The low energy region, with the many resonances and
threshold effects, is enhanced in the integral, and constitutes the
main difficulty of the method.
Alternative evaluations by lattice QCD are
continuously progressing, and are expected to become more and more competitive
in the near future \cite{Aoyama:2020ynm}.

A novel approach has been proposed in \cite{Calame:2015fva}, to determine the leading hadronic
contribution $a_{\mu}^{\rm HLO}$ from a measurement of the effective
electromagnetic coupling in the space-like region,
where the vacuum polarization is a smooth function. 
Its master equation is:
\begin{equation}\label{amu_xalpha}
        a_{\mu}^{\rm HLO} = 
         \frac{\alpha}{\pi} \int_0^1 dx \, (1-x) \,  \Delta \alpha_{\rm had} \! \left[ t(x) \right],
\end{equation}
where $\Delta\alpha_{\rm had}(t)$ is the hadronic contribution to the
running of the QED coupling, evaluated at 
\begin{equation}
        t(x)=-\frac{x^2m_\mu^2}{1-x} < 0,
\label{t}
\end{equation}
the space-like (negative) squared four-momentum transfer. By measuring
the running of the effective coupling: 
\begin{equation}\label{eq:alphaq2}
        \alpha(t) = \frac{\alpha(0)}{1-\Delta \alpha(t)},
\end{equation}
where $\alpha(0)=\alpha$ is the fine-structure constant,
the hadronic contribution $\Delta\alpha_{\rm had}(t)$ can be extracted
by subtracting from $\Delta \alpha(t)$ the purely leptonic part
$\Delta \alpha_{\rm lep} (t)$, which can be calculated to very high precision in QED.

To date there exist very few direct measurements of the running of $\alpha$
in the space-like region, the most precise one was obtained by the OPAL
experiment \cite{opal}, from small-angle Bhabha scattering, and reached
the sensitivity for the observation of the hadronic contribution.

Recently the MUonE experiment \cite{muone}
has been proposed, to measure the hadronic running of $\alpha(t)$
from $\mu e$ elastic scattering at low energy.
Here we will report the status of the project and few recent results.

\section{MUonE experimental proposal}
\begin{figure*}[!htbp]
\begin{center}
\includegraphics[scale=0.6]{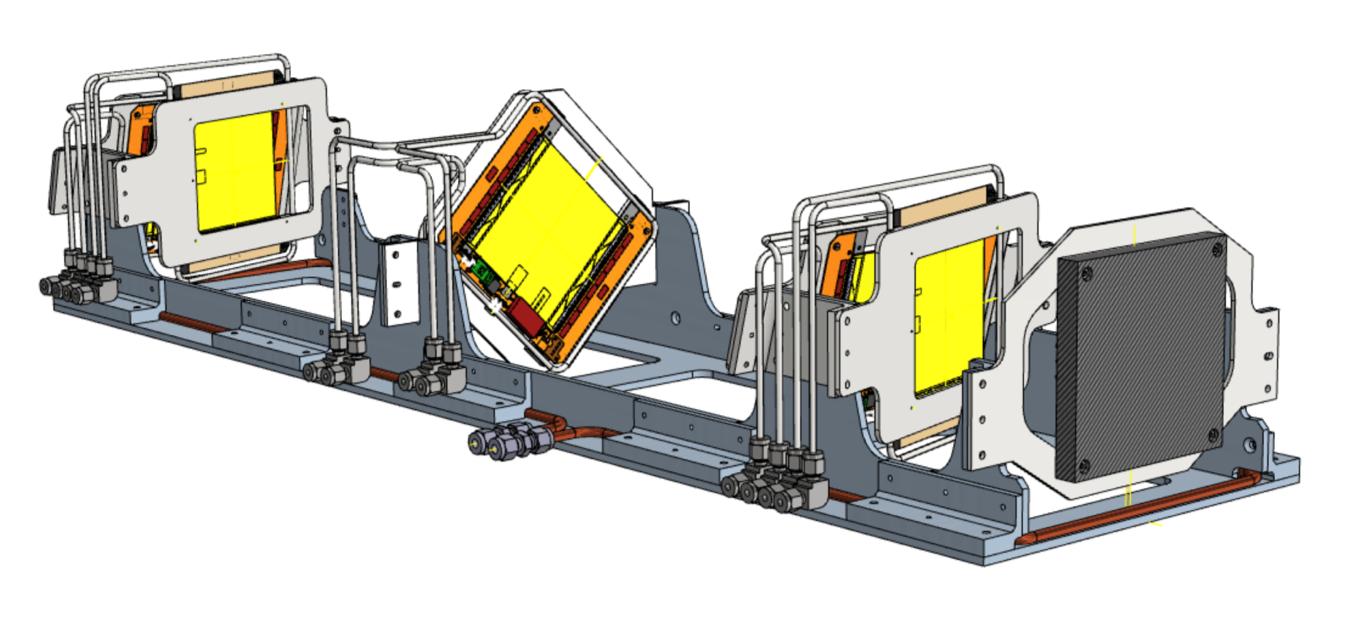}
\caption{CAD drawing of a MUonE tracking station.}
\label{MUonE-station-CAD}
\end{center}
\end{figure*}

MUonE aims at a very precise measurement of the shape of the
differential cross section of $\mu e$ elastic scattering, with the
CERN M2 muon beam ($E_\mu \sim 150$~GeV) off atomic electrons of a light Be or C target.
The hadronic contribution
$\Delta\alpha_{\rm had}(t)$ has a tiny variation across the probed
kinematical range $0 < x < 0.932$ (equivalent to $0 < -t < 0.143$~GeV$^2$), changing from a
vanishingly small value at low $x$ to about $10^{-3}$ at the peak
of the integrand in Eq.~\ref{amu_xalpha}, which occurs at $x = 0.914$
($t = -0.108 \mathrm{~GeV}^2$).
A competitive determination of $a_{\mu}^{\rm HLO}$ requires a
precision of ${\cal O}(10^{-2})$ in the measurement of the hadronic
running, which translates into an unprecedented precision of ${\cal
  O}(10^{-5})$ in the shape of the differential cross section.
To reach this impressive accuracy a modular detector is proposed, consisting
of a sequence of detection stations, each one made of a passive
element serving as target (1.5~cm Be plate) and active planes made of silicon microstrip
detectors for tracking, with length of 1 meter and transverse
dimensions of about 10 cm.
The current design of a MUonE tracking station is shown in
Fig.~\ref{MUonE-station-CAD}.
The complete layout foresees an array of 40 such stations, followed by an
electromagnetic calorimeter (ECAL) and a muon detector at the end.
This design could reach the target statistical sensitivity in three years
running time at the M2 beam, with an integrated luminosity of $1.5
\times 10^7 {\rm nb}^{-1}$. Clearly the ultimate challenge is keeping the systematics
at the same level as the statistical accuracy.

The basic tracking unit has been chosen to be the 2S module
developed for the upgrade of the CMS outer tracker
\cite{TRACKER_UPG_TDR}. It consists of two close-by planes of silicon
microstrips, providing track triggering capability at $40$~MHz, with inherent suppression of
background from single-layer hits or large-angle tracks.
It has a large active area of about $10\times10 \mathrm{~cm}^2$,
allowing to completely cover the relevant MUonE angular acceptance with a single module,
thus assuring the best uniformity. The position resolution $\sim$20
$\mu$m is adequate and can be further improved to $\sim$10$\mu$m
with a $\sim$15$^o$ tilt around the strip axis.
With respect to the LHC operation the main
difference for MUonE will be the asynchronous nature of the signals
from the $\mu e$ scattering events. In principle this can be managed with a
specific configuration of the front-ends but is a crucial issue
which will need thorough tests.

The MUonE project has been submitted to the CERN SPS Committee in June 2019
\cite{LoI}, obtaining positive recommendations and requests for
the necessary future milestones.

\section{Test Run 2021}
A first important milestone is a three-week Test Run allocated in the
last part of 2021 at the CERN M2 beam line, with full intensity muon
beam. The selected location is upstream of the COMPASS detector, after its
Beam Momentum Spectrometer (BMS).
The MUonE setup will consist of two tracking stations followed by the ECAL,
with an additional station (without target) upstream to track the incoming muons.

Each tracking station is a  1-meter long structure (Fig.~\ref{MUonE-station-CAD}),
with a target plate followed by three equally-spaced XY supermodules, each
one made of two 2S modules with orthogonal strips. The first and third
supermodules, measuring X and Y transverse coordinates, have tilted
detectors by $\sim$15$^o$ on the two orthogonal planes, to improve the
hit position resolution.
The middle supermodule is rotated by $45^o$
around the $z$ axis to solve the reconstruction ambiguities.
There are stringent requests on the mechanical stability of the
tracking stations, which has to be better than $10 \mu$m, in
particular on the longitudinal size.
Therefore the support structure is made of Invar (Fe-Ni alloy), which has a very low coefficient of thermal expansion, is
easy to machine and relatively cheap. 
A cooling system is also designed, in addition to an enclosure to
stabilize the room temperature within $1^o$-$2^o$.

The purpose of this Test Run is crucial. It should confirm the system
engineering, test the mechanical and thermal stability, the procedure
for the alignment (hardware and software), the readout chain and the
trigger strategy to identify and reconstruct $\mu e$ events.

Assuming to achieve these prime objectives and to have enough time
available for running, we have tried
to assess the physics potential of the chosen setup in the best-case scenario.
Considering the standard SPS efficiency and full beam intensity,
the two stations could potentially yield $\sim$1~pb$^{-1}/$day. 
By allowing for a reasonable time for detector commissioning and
considering possible DAQ inefficiencies we
could in principle integrate up to $\sim$5~pb$^{-1}$ of good data
during a first physics run, corresponding to $\sim$10$^9$ $\mu e$
scattering events with electron energy greater than 1~GeV.
Such a data sample would have enough sensitivity to measure the leptonic running
of $\alpha$ and, allowing some optimism, could even give initial sensitivity to
the hadronic running.

\begin{figure*}[!hbtp]
\begin{center}
\includegraphics[width=.5\textwidth]{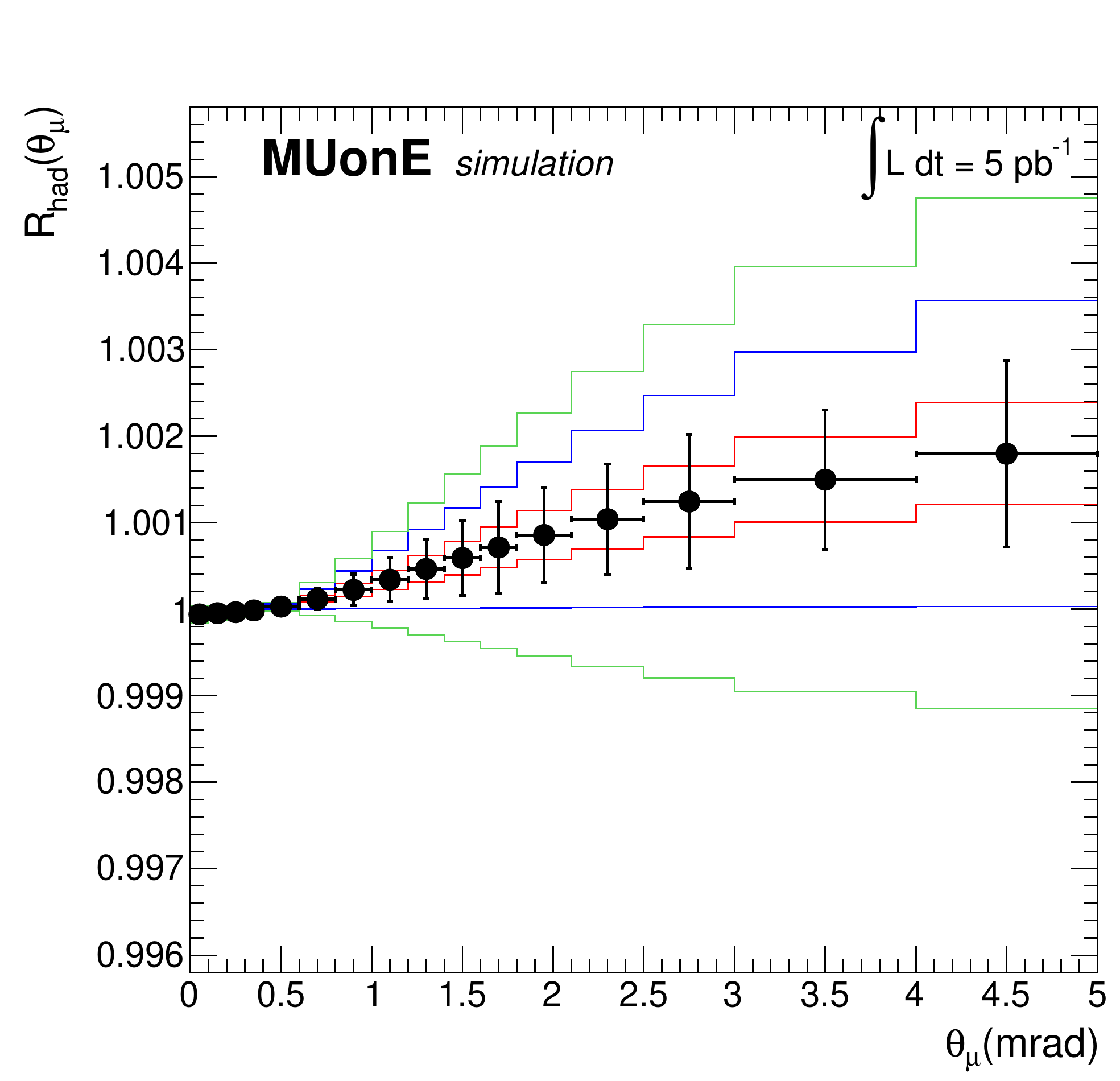}~\includegraphics[width=.5\textwidth]{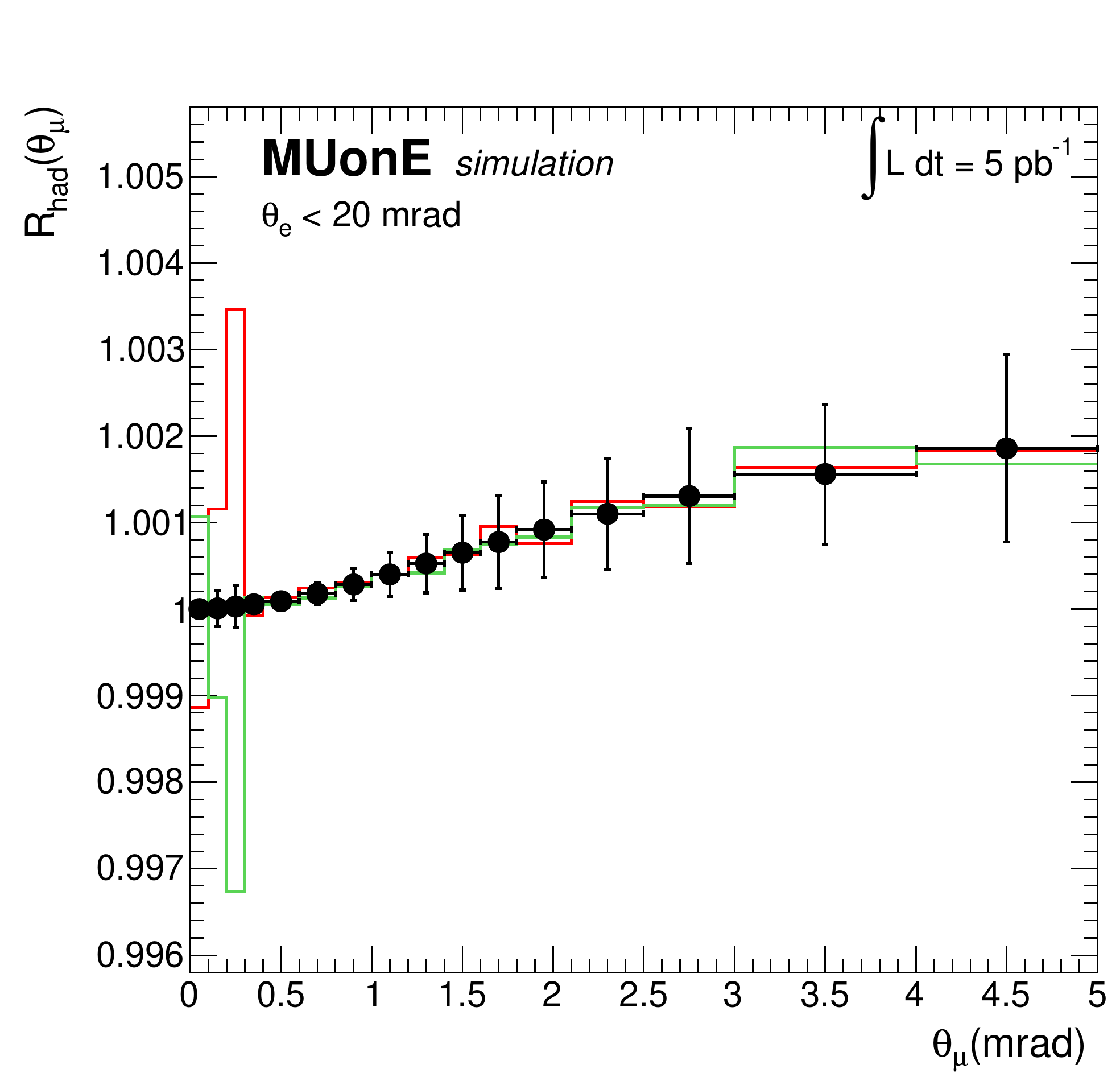}
\caption{Ratio $R_{\mathrm{had}}$ of the expected muon angular
  distribution and the prediction obtained with only leptonic
  running in $\alpha(t)$. The error bars correspond to the statistical
  uncertainties for an integrated luminosity of 5~pb$^{-1}$ assumed for the 2021 Test
  Run.
\emph{Left:} The histograms show the templates for few values
of the slope $K$.
\emph{Right:} The effect of a $+1\% (-1\%)$ systematic error on the assumed core
width of the MCS distribution is shown by the red (green) histogram. A
cut $\theta_e<20$~mrad is applied here to reduce the impact of this systematic.  
}
\label{TR21_fits}
\end{center}
\end{figure*}
\begin{figure*}[!htbp]
\begin{center}
\includegraphics[scale=0.4]{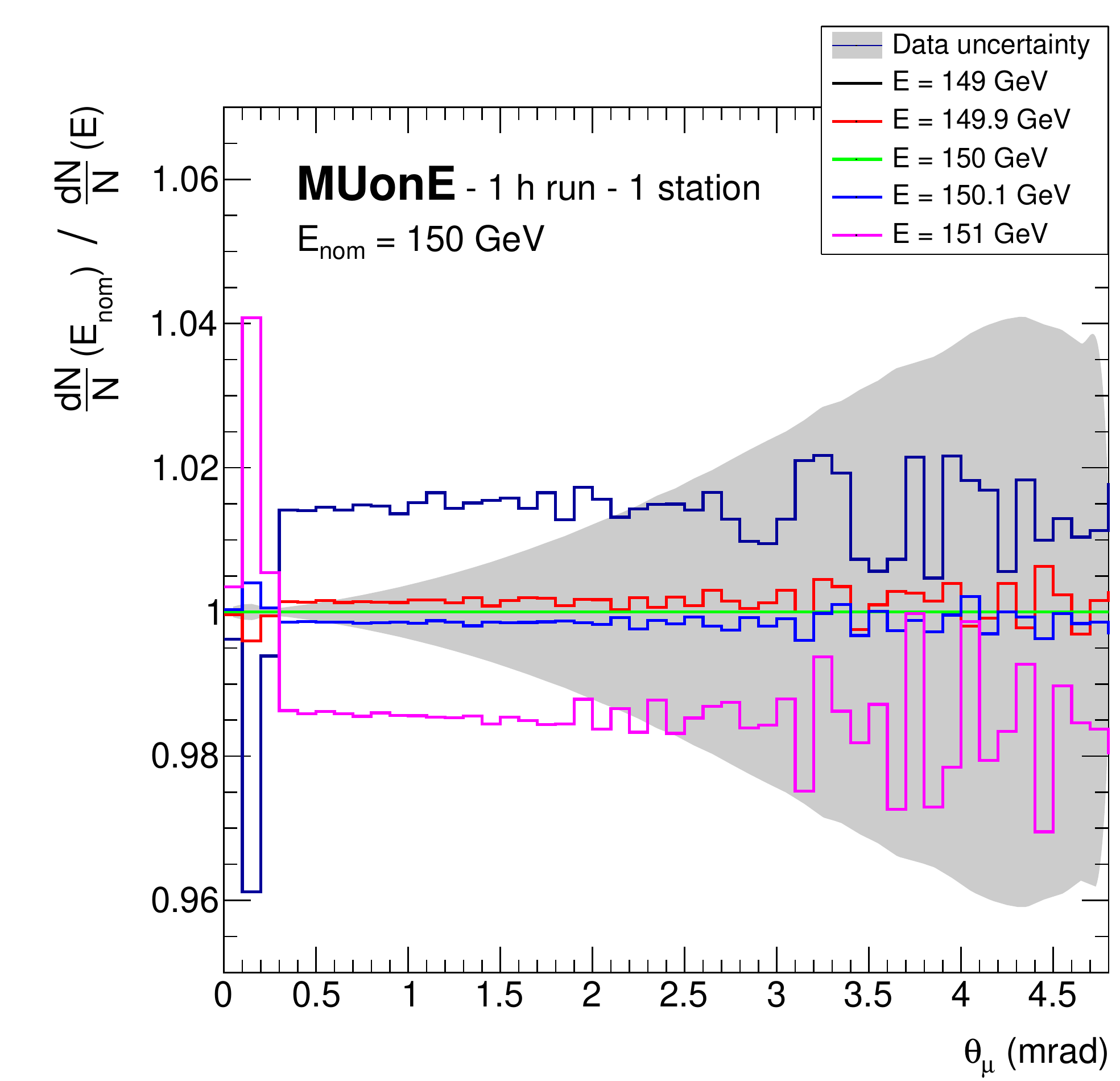}
\caption{Effect of a shift of $\pm (0.1$-$1.0)$~GeV in the average
  beam energy with respect to the nominal value $E_{\mathrm{nom}}=150$~GeV. The
  histograms show the expected distortions on the muon angular
  distribution, obtained from MC samples with variable energy. The grey band represents the statistical uncertainty
  corresponding to the expected data collected in one hour running
  time by one station.
}
\label{TR21_sys_escale}
\end{center}
\end{figure*}
The hadronic contribution to the running of $\alpha$ is most easily displayed by considering the ratio
$R_{\mathrm{had}}$ of the observed angular distributions with the theoretical
predictions evaluated for $\alpha(t)$ corresponding to only the
leptonic running.
Fig.~\ref{TR21_fits}-\emph{left} shows the
expectation for the muon angular distribution, obtained from a NLO MC
generator \cite{NLOgen} plus a fast simulation of
multiple Coulomb scattering (MCS) and the detector intrinsic resolution.
The extraction of the
hadronic contribution is carried out by a template fit method
\cite{LoI}, which here is simplified in a 1D fit as the limited
statistics allows to determine just a linear deviation on the shape.
The fit gives $K = 0.136 \pm 0.026$ as the slope of the observable
hadronic running at the purely statistical level.

It is important to assess the expected systematic errors, which will
have to be fully understood in the Test Run conditions to estimate
their impact in the full experiment. One of the most important
systematics is related to the estimated MCS, affecting in particular the low energy
electrons. This has been also studied in a dedicated beam test in 2017 \cite{TB2017}.
Fig.~\ref{TR21_fits}-\emph{right} 
shows the effect on the muon angular distribution of a flat $\pm1\%$
error on the MCS core width.
After a selection cut $\theta_e < 20 \mathrm{~mrad}$ on the electron angle and
by restricting the fit region to $\theta_\mu > 0.4 \mathrm{~mrad}$ one obtains:
$K = 0.137 \pm 0.033\emph{~(stat)~}^{+0.006}_{-0.004}\emph{~(syst)}$.

Another crucial systematic effect is related to the knowledge of the
average beam energy scale. This is known from the accelerator at a
level of about 1\%. The BMS spectrometer can measure individual incoming muons with
0.8\% resolution, and given the high muon beam intensity it can
provide an excellent monitor of time variations of the average scale.
However it cannot assess the systematic uncertainty of the average
energy scale, which has to be controlled by a physical process.
The kinematics of the elastic $\mu e$ scattering has been identified
as the useful method \cite{LoI}, in particular the average angle of
the two outgoing tracks, which does not need $\mu$-$e$ identification.
For illustration purpose Fig.~\ref{TR21_sys_escale} shows the effect
on the muon angular distribution
of a systematic error of $\pm (0.1$-$1.0)$~GeV on the
assumed average beam energy.
The expected distortion is compared to the statistical uncertainty corresponding to
one hour running time in one station.
It is clear that the energy calibration by the kinematical method would already outperform
the precision of the scale obtained from the accelerator.
The beam energy scale will be calibrated on each tracking station
independently, aiming at an ultimate precision for the final detector better than 3 MeV in less
than one week of run.

\section{Theoretical progress}
The MUonE challenging goal of a precision determination
of $a_{\mu}^{\rm HLO}$ depends also crucially on the availability of
very precise theoretical calculations for $\mu e$ scattering.
The full set of NLO QED and electroweak corrections was computed with
the development of a fully exclusive MC event generator \cite{NLOgen}.
The NNLO hadronic corrections have been recently computed in
\cite{Fael:2018dmz, Fael:2019nsf}.
The full NNLO QED corrections are not yet available, although several
important steps were taken in \cite{Mastrolia:2017pfy, DiVita:2018nnh, DiVita:2019lpl}.
For a state-of-the-art review see
\cite{muone-theory-initiative} and references therein.
On top of that a further milestone was recently achieved 
by the development of two independent fully exclusive MC codes, featuring the exact NNLO
photonic corrections on the leptonic legs, including all mass terms
\cite{MCNNLO_PV,MCNNLO_PSI}. Their results have been compared and are in very good agreement.
Resummation of leading terms in higher orders (Parton shower and YFS) matched to (N)NLO
will be also necessary and is underway.
Possible new physics effects in muon–electron collisions are expected
to lie below MUonE sensitivity \cite{Masiero:2020vxk, Dev:2020drf}.

\section{Next steps}
The MUonE Collaboration currently includes groups from CERN, China,
Germany, Greece, Italy, Poland, Russia, Switzerland, UK, and USA, in
addition to involved theorists from other countries.

On the experimental side an intense activity is going on for the preparation
of the Test Run, scheduled for the end of 2021.
This will be a proof of concept for the overall project and
is expected to accurately evaluate the experimental systematics, in order to assess the potential of the
proposed detector design to yield a competitive measurement.
If successful, a full proposal will be prepared including support from
the results of this run. The full detector construction could then take place
starting from 2022 with the prospect of a substantial running time during the LHC
Run3.


\end{document}